\documentclass[english,twocolumn]{revtex4}
\usepackage[L7x,T1]{fontenc}
\usepackage[latin9]{inputenc}
\usepackage{xcolor}
\usepackage{pdfcolmk}
\usepackage{array}
\usepackage{bm}
\usepackage{multirow}
\usepackage{amsmath}
\usepackage{graphicx}
\usepackage{esint}
\usepackage{bm}

\makeatother

\usepackage{babel}
\begin{document}
\global\long\def\icm{\mathrm{cm}^{-1}}
\global\long\def\vec#1{\bm{#1}}

\title{Distinctive character of electronic and vibrational coherences in
disordered molecular aggregates}

\author{Vytautas Butkus,$^{1,2}$ Donatas Zigmantas,$^{3}$ Darius Abramavicius,$^{1,4}$
Leonas Valkunas,$^{1,2}$ }

\email{leonas.valkunas@ff.vu.lt}

\address{$^{1}$Department of Theoretical Physics, Faculty of Physics, Vilnius
University, Sauletekio 9-III, 10222 Vilnius, Lithuania}

\address{$^{2}$Center for Physical Sciences and Technology, Gostauto 9, 01108
Vilnius, Lithuania}

\address{$^{3}$Department of Chemical Physics, Lund University, P.O. Box
124, 22100 Lund, Sweden}

\address{$^{4}$State Key Laboratory of Supramolecular Complexes, Jilin University,
2699 Qianjin Street, Changchun 130012, PR China}
\begin{abstract}
Coherent dynamics of coupled molecules are effectively characterized
by the two-dimensional (2D) electronic coherent spectroscopy. Depending
on the coupling between electronic and vibrational states, oscillating
signals of purely electronic, purely vibrational or mixed character
can be observed with the help of oscillation maps, constructed from
the time-resolved 2D spectra. The amplitude of the beatings caused
by the electronic coherence is heavily affected by the energetic disorder
and consequently the electronic coherences are quickly dephased. Beatings
with the vibrational character weakly depend on the disorder, assuring
their long-time survival. We show that detailed modeling of 2D spectroscopy
signals of molecular aggregates provides direct information on the
origin of the coherent beatings.
\end{abstract}
\maketitle

\section{Introduction}

Dynamic properties of electronic excitations in artificial and biological
molecular aggregates are determined by the intra- and intermolecular
interactions. Due to resonant intermolecular interactions, electronic
excitations of the aggregate form collective exciton states exhibiting
coherent relationship between molecular excited states \cite{PopeSwenberg1999,Amerongen2000}.
Interactions with intramolecular vibrations reflect the specificity
of molecular constituents of the aggregate and usually manifest themselves
as vibrational sub-bands in electronic absorption or fluorescence
spectra \cite{Fulton1964}. Depending on the relative strength of
these interactions, the induced excitations in molecular aggregates
may lead to a host of photoinduced dynamics: from coherent and incoherent
energy migration to reorganization of the surrounding environment
\cite{Sun2010,xu-yan-JCP2010,Gelzinis2011}.

To resolve the details of the excitation evolution time-resolved spectroscopy
techniques have been developed. Femtosecond coherent two-dimensional
(2D) spectroscopy is one of the most versatile techniques \cite{Mukamel2000,jonasARPC2003}.
2D spectroscopy has been already applied by considering the excitation
dynamics in various molecular aggregates, e.~g. photosynthetic pigment-protein
complexes, polymers, tubular aggregates, quantum dots \cite{Brixner2005,ZigmantasFlemingPNAS2006,engel-nat2007,Collini2009,Sperling2010,Turner2011}.
The inherent complexity of the 2D spectra was disclosed in all these
cases. For instance, the long-lasting oscillatory features, initially
attributed to electronic quantum coherences, have been resolved \cite{engel-nat2007,Collini2009,Westenhoff2012JACS,Panitchayangkoon2011}.
Possible vibronic/vibrational origin of some of these beats has been
proposed in a number of recent studies \cite{Christensson_JPCB2012,Butkus-Zigmantas-Abramavicius-Valkunas-CPL2012}.
Thus, it has been proved that coupling to multiple intramolecular
vibrational modes is essential to determine the spectral properties
of the most chromophore molecules resulting in complex behavior of
the 2D spectra \cite{Butkus-Zigmantas-Abramavicius-Valkunas-CPL2012,Jonas_PNAS2012,Mancal_JPC2012,Chin2013,Christensson_JPCB2012}.

Since the role of the coherence is considered to be an important factor
by defining the excitation dynamics in molecular aggregates the task
of distinguishing electronic and vibrational character of coherences
is essential and at the same time very challenging issue. To distinguish
between the origins of the spectral oscillating behavior a molecular
dimer (MD) as the simplest model of the vibronic molecular aggregate
is considered. This work is a natural extension of our previous comparative
study of coherent 2D spectra of vibrational monomer and electronically-coupled
dimer (ED) \cite{Butkus-Zigmantas-Abramavicius-Valkunas-CPL2012}.
Here we demonstrate that coherent oscillations of electronic and vibrational
character are mixed due to intermolecular interactions. However, they
exhibit distinct oscillation patterns and different behavior with
respect to the energetic disorder.

\section{Model}

As usual for resonant spectroscopy applications let us assume that
a constituent chromophore molecule of the aggregate is characterized
by two levels corresponding to electronic ground and excited states
and the electronic transitions are coupled to a single nuclear coordinate
of high-frequency intramolecular vibrational mode with frequency $\omega_{0}$.
This coupling can be effectively represented in terms of the displaced
harmonic oscillator model \cite{Amerongen2000}. The electronic excitation
results in the shift along the dimensionless coordinate $q$ of the
potential energy surface defined in the ground state as $V(q)=\omega_{0}q^{2}/2$.
Assuming the Heitler-London approximation \cite{PopeSwenberg1999}
the Hamiltonian of the dimer is given by 
\begin{align}
\hat{H} & =T+\left[V(q_{1})+V(q_{2})\right]|\mathrm{g}\rangle\langle\mathrm{g}|\nonumber \\
 & +\left[\varepsilon_{1}+V(q_{1}-d_{1})+V(q_{2})\right]|\mathrm{e}_{1}\rangle\langle\mathrm{e}_{1}|\nonumber \\
 & +\left[\varepsilon_{2}+V(q_{1})+V(q_{2}-d_{2})\right]|\mathrm{e}_{2}\rangle\langle\mathrm{e}_{2}|\nonumber \\
 & +J\left[|\mathrm{e}_{1}\rangle\langle\mathrm{e}_{2}|+|\mathrm{e}_{2}\rangle\langle\mathrm{e}_{1}|\right]\nonumber \\
 & +\left[\varepsilon_{1}+\varepsilon_{2}+V(q_{1}-d_{1})+V(q_{2}-d_{2})\right]|\mathrm{f}\rangle\langle\mathrm{f}|,\label{eq:Hcoord}
\end{align}
where $|\mathrm{g}\rangle$ is the electronic common ground state,
$|\mathrm{e}_{n}\rangle$ corresponds to the electronic excited state
of the $n$-th molecule of the dimer ($n=1,2$); $|\mathrm{f}\rangle$
is the doubly-excited state of the dimer, which corresponds to electronic
excitations of both monomers. Term $T$ represents the total kinetic
energy of the vibrational harmonic oscillators coupled to both molecules,
$d_{n}$ is the displacement value of the vibrational potential in
the electronically excited state, $\varepsilon_{n}$ is the energy
difference between the potential minima of $|\mathrm{g}\rangle$ and
$|\mathrm{e}_{n}\rangle$ states of the $n$-th monomer, $J$ is the
resonance interaction. Since $V(q_{n}-d_{n})-V(q_{n})=\lambda_{n}-\omega_{0}d_{n}q_{n}$
with $\lambda_{n}=\omega_{0}d_{n}^{2}/2$ being the so-called reorganization
energy, this difference in the potential energies determines the coupling
of the electronic excitation to molecular vibrations. The latter is
usually characterized by the dimensionless Huang-Rhys factor $s_{n}=d_{n}^{2}/2$.

\begin{figure}
\includegraphics{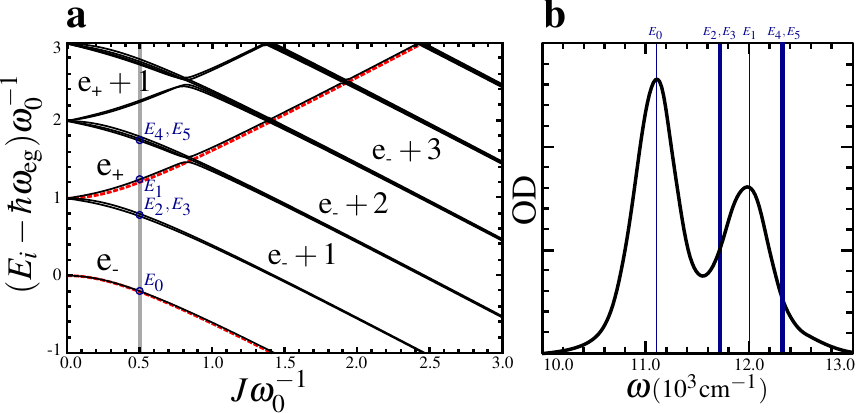}

\caption{\label{fig:Dependence-of-the}Dependence of the energies of the singly-excited
states of the molecular dimer on the resonant coupling (a, black solid
lines) in the case of small Huang-Rhys factors $s_{1}=s_{2}=0.05$.
A clear separation of electronic excited states (denoted as ${\rm e}_{-}$
and ${\rm e}_{+}$, respectively) and vibronic states (denoted as
${\rm e}_{\pm}+m$, where $m$ -- the number of vibrational quanta)
is observerd. The energies of a purely electronic dimer $E_{e_{\pm}}$
are represented by the red dashed lines. (b) The absorption spectrum
of the molecular vibronic dimer with coupling $J=0.5\omega_{0}$ is
also shown, which corresponds to the energy states indicated by the
vertical gray line in (a). Positions of the transitions contributing
to the absorption spectrum are indicated by vertical lines.}
\end{figure}

In the case of no displacement ($s_{1}=s_{2}=0$) the electronic excitations
are not coupled to molecular vibrations and the electronic part of
the Hamiltonian is easily diagonalized giving analytical expressions
of the singly-excited state energies of the electronic dimer, $E_{e_{\pm}}=\omega_{{\rm eg}}\pm\Delta E_{{\rm ED}}$,
where $\omega_{{\rm eg}}=\frac{1}{2}(\varepsilon_{1}+\varepsilon_{2})$
and $\Delta E_{{\rm ED}}=\frac{1}{2}\sqrt{(\varepsilon_{2}-\varepsilon_{1})^{2}+4J^{2}}$.
Dependence of the exciton energies of the electronic dimer on the
coupling is shown in Fig.~\ref{fig:Dependence-of-the}a by the red
dashed lines. The remaining vibrational Hamiltonian corresponds to
the harmonic oscillator which can be easily quantized.

The ground-state Hamiltonian in Eq.~(\ref{eq:Hcoord}) is diagonal
by definition independently on the value of $s_{n}$ and contains
a ladder of vibrational states with energies spaced by $\omega_{0}$.
The singly-excited state block of the Hamiltonian is not diagonal
in this basis and depends on the value of $s_{n}$. After numerical
diagonalization, if a sufficient number of vibrational states $\nu_{{\rm vib}}$
is taken into account, the excitation energies are found approximately
at $E_{e_{-}}+\omega_{0}m$ and $E_{e_{+}}+\omega_{0}m$ ($m=0,1,\ldots$)
and virtually constitute two ladders of equally-spaced states, denoted
by $e_{+}+m$ and $e_{-}+m$ in Fig.~\ref{fig:Dependence-of-the}a.
In the energy spectrum, the avoided crossing regions, where the electronic
splitting matches vibrational quantum, emerge at particular values
of the parameters when the most pronounced mixing of state character
occurs \cite{Polyutov2012}. For the larger Huang-Rhys factors ($s=0.05$
is used in Fig.~\ref{fig:Dependence-of-the}a) the deviation of the
vibrational excitation energies from the harmonic ladder structure
becomes significant, especially in the vicinity of the avoided crossing
regions.

We denote the electronic ground state as the state where both molecules
are in their ground states with arbitrary vibrational excitations,
$|{\rm g}_{i}{\rm g}_{j}\rangle$. Index $i$ here indicates the $i$-th
vibrational excitation of the first molecule and $j$ denotes the
vibrational excitation of the second molecule. $|{\rm e}_{1}\rangle=|{\rm e}_{i}{\rm g}_{j}\rangle$
denotes the state of the first molecule being in the vibronic excited
state maintaining the $i$-th vibration quantum and the second molecule
being in the $j$-th vibrational ground state; analogously, $|{\rm e}_{2}\rangle=|{\rm g}_{i}{\rm e}_{j}\rangle$
corresponds to the state where the second molecule is vibronically
excited. Doubly-excited electronic states $|\mathrm{f}\rangle=|{\rm e}_{i}{\rm e}_{j}\rangle$
are constructed in a similar way. Thus, the site basis of our model
is complete with respect to vibronic states since one-particle vibronic
states (such as $|{\rm e}_{i}{\rm g}_{0}\rangle$) and two-particle
vibronic states ($|{\rm e}_{i}{\rm g}_{j}\rangle,$ $j\ne0$) are
included. In terms of these definitions the vibronic set of eigenstates
for the singly- and doubly-excited electronic states (state vectors
$|u\rangle$ and $|v\rangle$, correspondingly) can be obtained by
using the relevant transformations $|u\rangle=\sum_{ij}\left(\phi_{u,{\rm e}_{i}{\rm g}_{j}}^{(1)}|{\rm e}_{i}{\rm g}_{j}\rangle+\phi_{u,{\rm g}_{i}{\rm e}_{j}}^{(2)}|{\rm g}_{i}{\rm e}_{j}\rangle\right)$
and $|v\rangle=\sum_{ij}\Phi_{v,{\rm e}_{i}{\rm e}_{j}}|{\rm e}_{i}{\rm e}_{j}\rangle$.
The corresponding transformation coefficients $\phi^{(n)}$ and $\Phi$
are acquired from the diagonalization of the singly-excited and doubly-excited
blocks of the Hamiltonian~(\ref{eq:Hcoord}) and provide us with
the information about the eigenstate composition which allows the
estimation of the amount of mixing between vibrational and electronic
states.

The transition dipole moments between the ground and singly-excited
states as well as between singly- and doubly-excited states in the
vibronic eigenstate basis are then given by
\begin{equation}
\vec{\mu}_{u,{\rm g}_{i}{\rm g}_{j}}=\vec{\mu}_{1}\phi_{u,{\rm e}_{i}{\rm g}_{j}}^{(1)}+\vec{\mu}_{2}\phi_{u,{\rm g}_{i}{\rm e}_{j}}^{(2)}
\end{equation}
and

\begin{equation}
\vec{\mu}_{vu}=\vec{\mu}_{2}\sum_{ijlm}\phi_{u,{\rm e}_{l}{\rm g}_{m}}^{(1)}\Phi_{v,{\rm e}_{i}{\rm e}_{j}}+\vec{\mu}_{1}\sum_{ijlm}\phi_{u,{\rm g}_{l}{\rm e}_{m}}^{(2)}\Phi_{v,{\rm e}_{i}{\rm e}_{j}},
\end{equation}
where $\vec{\mu}_{1}$ and $\vec{\mu}_{2}$ are the transition dipole
moments of the monomers.

We will consider 2D signals of the molecular dimer, which are calculated
using the perturbative system-response function theory in the vibronic
eigenstate basis using the impulsive limit of the laser pulses, i.e.
assuming that the laser pulse spectrum covers all frequencies. More
information regarding the details of calculations can by found elsewhere
\cite{Abramavicius2010,Abramavicius2007}. For the sake of simplicity,
we assume the pure dephasing as the only mechanism responsible for
the homogeneous lineshape formation. The energetic disorder (uncorrelated
fluctuations of site energies $\varepsilon_{n}$) is considered to
be responsible for the inhomogeneous broadening that is taken into
account by averaging over ensemble (1000 realizations of independent
simulations) with the Gaussian distribution with standard deviation
$\sigma_{{\rm D}}$ of excitation energies for every molecule.

\section{Results}

We consider the molecular dimer with Huang-Rhys factors equal for
both monomers, $s\equiv s_{1}=s_{2}$. We set vibrational frequency
$\omega_{0}$ as the reference parameter and choose the resonant coupling
strength $J=-\omega_{0}/2$. We also choose that electronic site energies
are separated by the same value as vibrational frequency, $\varepsilon_{2}-\varepsilon_{1}=\omega_{0}$.
As values typical for aggregates, we consider $\omega_{0}=600\,\icm$
and $s=0.05$, thus resulting in $\Delta E_{{\rm MD}}$ $\approx867\,\icm$,
which is approximately equal to the excitonic splitting of the electronic
dimer $\Delta E_{{\rm ED}}$. Hence, the excitonic splitting is off-resonant
with respect to the vibronic resonance (compare ${\rm e}_{+}$, ${\rm e_{-}}$
and ${\rm e}_{-}+1$ levels Fig.~\ref{fig:Dependence-of-the}a).
We assume that the strengths of the transition dipole moments of the
monomers are equal and constitute the inter-dipole angle $\varphi=\frac{2\pi}{5}$.
The dephasing rate determining the homogeneous linewidth is set to
$\gamma=\omega_{0}/6$ and central absorption frequency is chosen
$\omega_{{\rm eg}}=11500\,{\rm cm}^{-1}$. For numerical calculations
we choose $\nu_{{\rm vib}}=8$; calculations with more vibrational
states show no evident changes in any simulated spectroscopic signals
for the used set of parameters. The deviation from a harmonic ladder
of vibronic sates appears, but is insignificant for the model considered
here, as the singly-excited state energy departure from the ideal
harmonic progression is less than 5\% and the chosen parameters ensure
that we are away from the avoided-crossing regions (see Fig.~\ref{fig:Dependence-of-the}a).

The electronic interaction with the high-frequency vibrational mode
is almost indistinguishable in the absorption spectrum for the chosen
small value of the Huang-Rhys factor as the only evidence of vibronic
content is a sole weak shoulder in absorption at $\sim12600\,{\rm cm}^{-1}$
(Fig.~\ref{fig:Dependence-of-the}b). For the larger Huang-Rhys factor
the stronger vibrational progression would evidently appear in the
spectrum. However, consideration of such an aggregate instead would
not change conclusions of the further discussion appreciably.

Transition amplitudes $\mu_{u,{\rm g}_{i}{\rm g}_{j}}$ between the
ground state manifold of the vibrational states and the manifold corresponding
to the singly-excited states signify possible interaction configurations.
The \char`\"{}vibronic content\char`\"{} in a specific electronic
transition can be quantified by the transformation coefficients combined
as $\chi_{u}\equiv\left(\phi_{u,{\rm g}_{0}{\rm g}_{0}}^{(1)}\right)^{2}+\left(\phi_{u,{\rm g_{0}}{\rm g}_{0}}^{(2)}\right)^{2}$.
The maximum value of this quantity ($\chi_{u}=1$) indicates the pure
electronic character of the state, while $\chi_{u}=0$ reflects that
the corresponding transition is vibronic. Transitions originating
from the zero-vibrational state, i.~e. $|{\rm g}_{0}{\rm g}_{0}\rangle$
are described in Table~\ref{tab:dipoles}. The two most significant
transitions correspond to electronic-only transitions with $\chi\geq0.9$,
while the other transitions are of dominant vibrational character
($\chi<0.1$).

\begin{table*}
\begin{tabular}{ccccccccccc}
\multirow{2}{*}{$u$} & \multirow{2}{*}{$E_{u}-E_{0}$, cm$^{-1}$} & \multirow{2}{*}{$\mu_{u,g_{0}g_{0}}$} & \multirow{2}{*}{$\chi_{u}$} & \multirow{2}{*}{$\varphi_{u}$, deg} & \multicolumn{6}{c}{States in site basis}\tabularnewline
\cline{6-11} 
 &  &  &  &  & $|e_{0}g_{0}\rangle$ & $|g_{0}e_{0}\rangle$ & $|e_{1}g_{0}\rangle$ & $|g_{0}e_{1}\rangle$ & $|e_{0}g_{1}\rangle$ & $|g_{1}e_{0}\rangle$\tabularnewline
\hline 
0 & 0 & 1.08 & 0.96 & 19.0 & \textbf{0.82} & \textbf{0.14} & 0.03 & 0.00 & 0.00 & 0.00\tabularnewline
1 & 867 & 0.84 & 0.90 & -81.8 & \textbf{0.15} & \textbf{0.76} & 0.02 & 0.01 & 0.05 & 0.01\tabularnewline
2 & 581 & 0.26 & 0.06 & 69.3 & 0.00 & 0.06 & \textbf{0.37} & 0.09 & \textbf{0.37} & 0.06\tabularnewline
3 & 600 & 0.17 & 0.02 & 19.0 & 0.02 & 0.00 & \textbf{0.38} & 0.07 & \textbf{0.41} & 0.07\tabularnewline
4 & 1467 & 0.13 & 0.02 & -81.3 & 0.00 & 0.02 & 0.09 & \textbf{0.36} & 0.06 & \textbf{0.36}\tabularnewline
5 & 1480 & 0.11 & 0.01 & 69.8 & 0.00 & 0.01 & 0.08 & \textbf{0.33} & 0.08 & \textbf{0.36}\tabularnewline
\hline 
\end{tabular}

\caption{\label{tab:dipoles}Mixed character of the lowest vibronic eigenstates
of the molecular dimer with the Huang-Rhys factor $s=0.05$. Value
of $\chi_{u}=1$ signifies a completely electronic character of the
exciton state, while $\chi_{u}=0$ corresponds to the vibrational
character. $\varphi_{u}$ denotes the angle of corresponding transition
dipole vector with respect to the electronic transition dipole of
the first monomer. $E_{u}-E_{0}$ correponds to the energy gap between
vibronic sates and $\mu_{u,{\rm g}_{0}{\rm g}_{0}}$-- to the transition
dipole moment. }
\end{table*}

\begin{figure}
\includegraphics{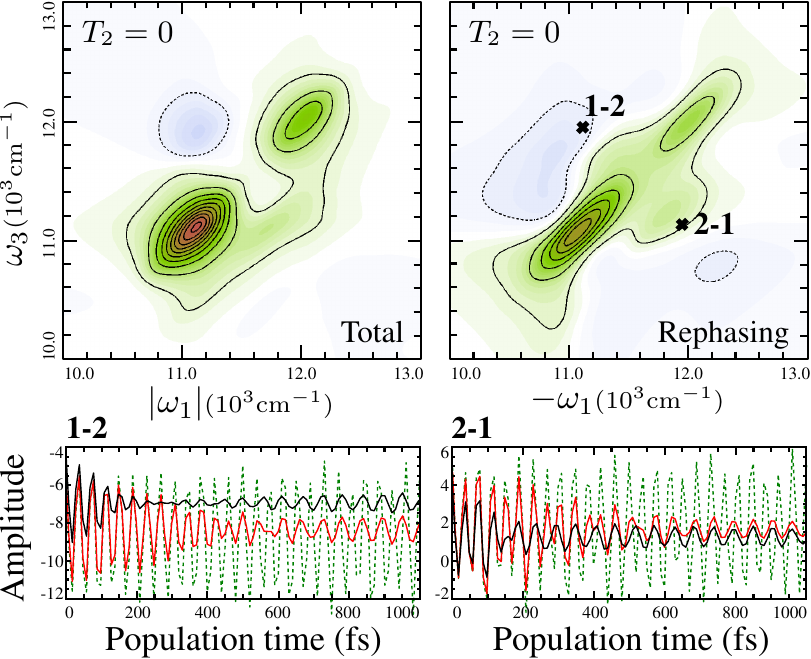}

\caption{\label{fig:Time-dependence-of}The total (a sum of rephasing and non-rephasing)
and only rephasing 2D spectra of the molecular dimer at the population
time $T_{2}=0$, with Gaussian disorder $\sigma_{{\rm D}}=200$~${\rm cm}^{-1}$
(upper panels). Solid (dashed) contour lines for positive (negative)
signals are drawn in 10\% intervals. Time dependences of the amplitudes
of the upper (1-2) and lower (2-1) cross-peaks of the rephasing signal
are shown in the lower panels (left and right, respectively). Green
-- no disorder, red -- $\sigma_{{\rm D}}=20\,\icm$, black -- $\sigma_{{\rm D}}=50\,\icm$. }
\end{figure}

\begin{figure}
\includegraphics{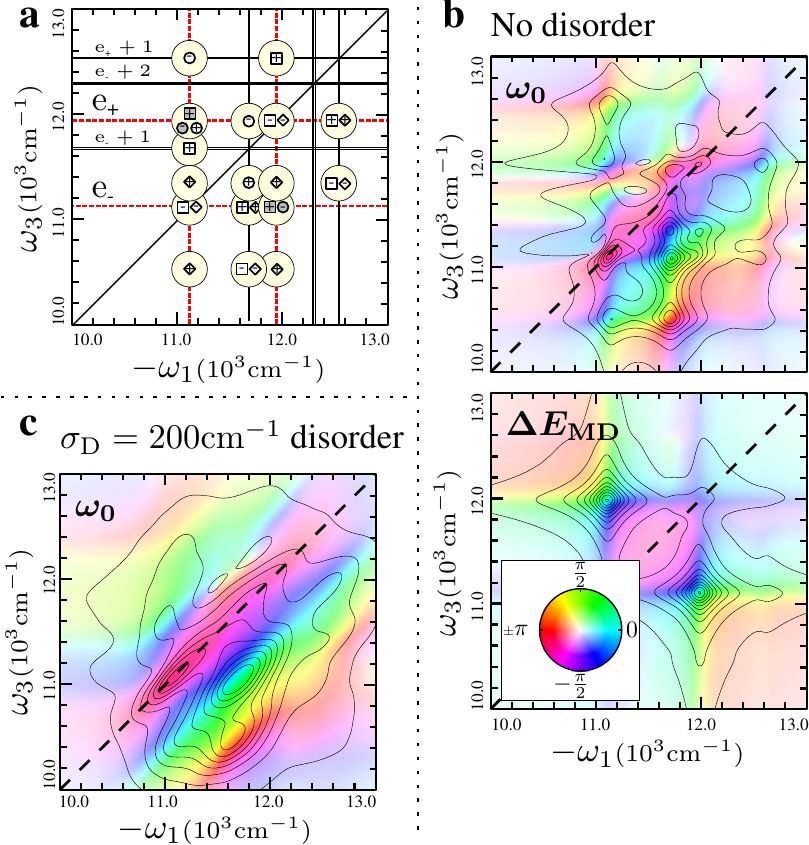}

\caption{\label{fig:diagrams_smallHR}(a) Arrangement of the the most significant
(90\% of the total amplitude) oscillatory contributions (diamonds
denote GSB, squares -- SE and circles -- ESA) in the 2D spectrum of
the molecular dimer. The oscillations are at vibrational frequency
$\omega_{0}$ (open symbols) or electronic frequency $\Delta E_{{\rm MD}}$
(full gray symbols) and the plus/minus sign at each peak denotes the
phase of the oscillation. (b) The Fourier maps are represented by
the amplitude and phase of oscillations as contour lines and peak
color, respectively, at frequencies $\omega_{0}$ and $\Delta E_{{\rm MD}}$
when $\sigma_{{\rm D}}=0$. (c) The oscillation map at frequency $\omega_{0}$
when $\sigma_{{\rm D}}=200\,\icm$. The map at $\Delta E_{{\rm MD}}$
frequency is not shown since it has a negligible amplitude. }
\end{figure}

The real parts of rephasing and total 2D spectra contain three positive
features: two diagonal peaks and a clearly distinguishable cross-peak
(2-1) below the diagonal reflecting the coherent resonance coupling
between the molecules (Fig.~\ref{fig:Time-dependence-of}). The higher
cross-peak (1-2) is less visible due to the overlap with the excited
state absorption (ESA) contribution. Apparent simple structure of
the spectra the complicated pattern of various overlapping contributions.
They can be sorted into: i) \emph{stationary contributions}, denoting
the excitation pathways, where the aggregate is in a population state
during delay time $T_{2}$; ii) \emph{oscillating contributions},
reflecting the coherent electronic or vibronic excitation dynamics
in the excited state or vibrational dynamics in the ground state.
The multitude of the later ones in case of the molecular dimer system
is illustrated in Fig.~\ref{fig:diagrams_smallHR}a. Temporal dynamics
of oscillating contributions can be represented by the time dependent
traces of certain spectral regions. Such traces of upper (1-2) and
lower (2-1) cross-peaks of the rephasing signal are presented in the
lower panel of Fig. \ref{fig:Time-dependence-of}. The oscillations
can also be characterized by studying the so-called oscillation Fourier
maps which are constructed by applying the Fourier transform to the
real part of the rephasing spectrum $S_{{\rm R}}(\omega_{3},T_{2},\omega_{1})$
with respect to the delay time $T_{2}$ \cite{Turner2012,Calhoun2009},
\begin{equation}
A(\omega_{3},\omega_{2},\omega_{1})=\intop_{0}^{\infty}{\rm d}T_{2}e^{{\rm i}\omega_{2}T_{2}}{\rm Re}S_{{\rm R}}(\omega_{3},T_{2},\omega_{1}).
\end{equation}
This allows us to directly identify the phase and amplitude of oscillations
in the different spectral regions. Such oscillation maps of the 2D
spectra at frequencies corresponding to $\omega_{2}=\omega_{0}$ and
$\omega_{2}=\Delta E_{{\rm MD}}$ are presented in figures~\ref{fig:diagrams_smallHR}b
and \ref{fig:diagrams_smallHR}c in case of no disorder ($\sigma_{{\rm D}}=0$)
and substantial disorder ($\sigma_{{\rm D}}=200\,\icm$), respectively.
The oscillation map at $\omega_{2}=\Delta E_{{\rm MD}}$ involves
two states with the most pronounced electronic character (see Table~\ref{tab:dipoles})
and is the same as the purely electronic coherence map of an excitonic
dimer \cite{Butkus-Zigmantas-Abramavicius-Valkunas-CPL2012}, i.e.
oscillations in the rephasing part appear only in the cross-peaks.
As we find in the disordered system ($\sigma_{{\rm D}}=200\,\icm$)
the oscillating patterns are completely dominated by the vibrational
frequency $\omega_{0}$, while the oscillations with the electronic
gap frequency $\Delta E_{{\rm MD}}$ are at least 20 times weaker
and, thus, their contribution is negligible. The oscillation map of
the disordered vibronic dimer is highly asymmetric, which is mostly
due to the ground state bleaching (GSB) contribution (Fig.~\ref{fig:diagrams_smallHR}a).

It is known that the phase of spectral oscillations of the electronic-only
systems is 0 at the maxima of the peaks and changes continuously when
going away from the peak (rises in direction outwards the diagonal).
For the monomer coupled to a single mode of high-frequency vibration,
the phase of either $0$ or $\pi$ can be obtained depending on the
coupling strength \cite{Butkus-Abramavicius-Valkunas-JCP2012}. However,
in congested spectra any phase relationships can be obtained if the
spectral overlap is substantial. Especially for the mixed system,
when both resonance coupling and vibronic interactions are taken into
account. For example, it can be seen in the oscillation map in Fig.~\ref{fig:diagrams_smallHR}c
that at the center of the lower diagonal peak the phase is detuned
by $\pi/3$ from the value of $\pi$, which would be expected for
an isolated oscillating peak \cite{Butkus-Abramavicius-Valkunas-JCP2012,Jonas_PNAS2012}.

\section{Discussion}

The parameter set used in calculations describes a very general electronic-vibronic
system. The chosen values $\left|J\right|=\omega_{0}/2$ do not correspond
to any special case since the monomer energies remain different, $\varepsilon_{2}\ne\varepsilon_{1}$.
The chosen absolute values of parameters are typical for many molecular
aggregates of interest, including the photosynthetic pigment-protein
complexes. Similar parameters have already been used by investigating
the vibronic transition dipole moment borrowing and coherence enhancement
effects in molecular dimer systems \cite{Polyutov2012,Chenu2013}.
The introduced model captures both limiting cases of purely vibrational
and purely electronic model systems that are usually considered separately.
Indeed, by assuming either the Huang-Rhys factors or the resonance
interaction to be zero both limiting cases can be obtained. Two prominent
frequencies are present in the simulated oscillatory dynamics of the
2D spectrum of the vibronic molecular system. Evidently, one frequency
corresponds to the electronic energy gap $\Delta E_{{\rm MD}}$, while
the other is equal to the vibrational frequency $\omega_{0}$. These
coherences are between the states of mixed electronic-only and electronic-vibronic
character, respectively (Table~\ref{tab:dipoles}). It is noteworthy
to mention, that a very similar amount of mixing is obtained using
even simpler description with only one-particle vibronic states and
$\nu_{{\rm vib}}=2$ included \cite{Polyutov2012}. The mixed character
might indirectly appear more signifficant if the diagonal disorder
is comparable to the gap between states of different character. Then,
the states of different character become overlapping leading to more
mixing \cite{Chenu2013}. However, the distribution of the transition
frequencies not the amount of mixing of particular state defines the
oscillations. Thus, beatings with rather electronic and rather vibrational
character still result in distinct oscillation maps and therefore
the origin of the oscillation can be identified.

\begin{figure}
\includegraphics{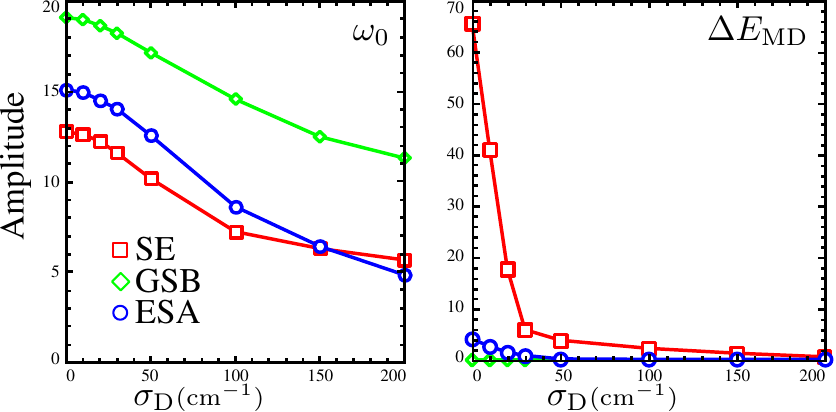}

\caption{\label{fig:amplitudes_diagrams}Maximal amplitude of the oscillation,
taken from the oscillation maps as a function of disorder for GSB
(diamonds), SE (squares) and ESA (circles) contributions at different
beating frequencies ($\omega_{0}$ and $\Delta E_{{\rm MD}}$). }
\end{figure}

Strength of oscillations strongly depends on system inhomogeneity.
From simulations, this can clearly be seen in the lower panel of Fig.~\ref{fig:Time-dependence-of},
where time dependences of the peak value for different values of the
Gaussian disorder ($\sigma_{{\rm D}}=0$, $20$ and $50\,{\rm cm}^{-1}$)
are presented. The initial intensive oscillations with the $\Delta E_{{\rm MD}}$
frequency decay rapidly when the disorder is increased and the only
dynamics observed at longer delay times correspond to the $\omega_{0}$
beats. Strengths of the oscillations at different frequencies are
represented by the corresponding values of the amplitude in the oscillation
maps. Therefore, the value of the amplitude maximum of the oscillation
map is used to evaluate the strength of the corresponding oscillation.
The amplitude of electronic oscillations decays sharply with the disorder,
while dependence of the amplitude of vibronic/vibrational coherences
is much more flat (the amplitudes for the stimulated emission (SE),
excited state absorption (ESA) and ground state bleaching (GSB) contributions
are presented in Fig.~\ref{fig:amplitudes_diagrams}). When disorder
is absent, signatures of electronic coherences mostly coming from
the SE contribution are at least 5 times larger than those of vibrational
character, while for $\sigma_{{\rm D}}>200\,{\rm cm}^{-1}$ the amplitude
of electronic coherences is negligible (the corresponding oscillation
map for $\sigma_{{\rm D}}=200\,{\rm cm}^{-1}$ cannot be resolved).

Separation of coherences of electronic and vibrational character is
very significant as our results demonstrate that besides vibrational
beats the electronic beats could be in principle observed and distinguished
for the weakly disordered systems. However, the electronic beats rapidly
decay in time in a Gaussian fashion (for the Gaussian disorder) as
$\sigma_{{\rm D}}^{-1}$. Whereas vibrational beats will prevail for
longer times.

Note that in the Hamiltonian used (Eq.~\ref{eq:Hcoord}) no transformation
of canonical variables into symmetric and antisymmetric components
is performed that would allow us to decouple the Schrdinger equations
of the excited and ground states. Therefore, presented model contains
states which are not separated to diabatic or adiabatic ones as has
been done in the theoretical work of Jonas and coworkers addressing
coherences in FMO \cite{Jonas_PNAS2012}. Also, since we show the
separation of beats of electronic and vibrational character for arbitrary
system parameters, the resonant condition of the mixed character coherences
(for example, when electronic energy gap or the chromophore energy
difference is equal to the vibrational frequency) is just a special
case of our model. It supports the assumptions about coherences in
FMO where such resonant conditions are met -- electronic coherences
at initial times could be of the same order as those of the vibronic
origin due to disorder of approximately $25$~cm$^{-1}$; electronic
beats will decay in a short time (\textasciitilde{}200 fs) while vibrational
beats would persist over the long time. For higly disordered and uncorrelated
systems electronic coherences are not likely to be significant at
all and the observed coherent dynamics are due to the ground-state
vibrations.

It is also important to consider transition dipole moment orientations
$\varphi_{u}$ listed in Table~\ref{tab:dipoles}. The orientations
of the transitions to the ``mixed'' states are different for each
state and are also different from the transition dipole moments of
monomer transitions. This implies that coherences involving arbitrary
states (of electronic or vibrational character) might survive the
measurements with polarization schemes, devised to suppress all but
electronic coherence signals \cite{Westenhoff2012JACS,Schlau-Cohen2009}.
On the other hand, these polarization schemes can then be used to
distinguish between purely vibrational (localized on one molecule)
and mixed origin coherences.

Results presented here are related to the assumption that the vibrational
frequencies are not affected by inhomogeneities, which induce the
disorder of electronic transition energy. This is often the case as
vibrational resonances are less sensitive to the electrostatic configuration
of the environment than the delocalized electronic wavefunctions.
Vibrational coherences, hence, decay on a timescale of vibrational
dephasing, which usually is in the order of a picosecond. It should
be also noted that when vibrational frequency $\omega_{0}$ matches
the electronic energy gap $\Delta E_{{\rm MD}}$, the quantum mechanical
mixing complicates the whole picture. Then damping of the resulting
electronic-vibrational coherence will define the decay timescale of
the oscillations in the spectra.

\section{Conclusions}

Coherent dynamics of coupled molecules can be effectively sorted out
by the modeling of 2D spectra. Coupling between the multitude of vibrational
states on different molecules results in a manifold of mixed states.
When vibrational frequency and the electronic energy gap are off resonance
this manifold constitutes two ladders of equally-spaced and well-resolved
states. Two types of beats having either electronic or vibrational
character can then be distinguished by the use of oscillation maps,
constructed from the sequences of time-resolved 2D spectra. 

Inhomogeneous disorder effects the coherences of different state character
differently. The amplitude of the electronic character beatings, caused
by the coherences in excited states, is dramatically reduced by the
disorder and consequently electronic coherences are quickly dephased.
Vibrational character beatings stem from the ground and excited state
contributions and depend weakly on the disorder, assuring their long-time
survival.

\section*{Acknowledgments}

This research was partially funded by the European Social Fund under
the Global grant measure. V.~B. acknowledges support by project \char`\"{}Promotion
of Student Scientific Activities\char`\"{} (VP1-3.1-MM-01-V-02-003)
from the Research Council of Lithuania. D.~Z was supported by the
Swedish Research Council.


\begin{thebibliography}{10}

\bibitem{PopeSwenberg1999}
M.~Pope and C.~E. Swenberg.
\newblock {\em Electronic processes in organic crystals and polymers}.
\newblock Oxford University Press, New York/Oxford, 2 edition, 1999.

\bibitem{Amerongen2000}
H.~van Amerongen, L.~Valkunas, and R.~van Grondelle.
\newblock {\em Photosynthetic Excitons}.
\newblock World Scientific Co., Singapore, 2000.

\bibitem{Fulton1964}
Robert~L. Fulton and Martin Gouterman.
\newblock Vibronic coupling. {II}. {S}pectra of dimers.
\newblock {\em {J}. {C}hem. {P}hys.}, 41(8):2280--2286, 1964.

\bibitem{Sun2010}
Jin Sun, Bin Luo, and Yang Zhao.
\newblock Dynamics of a one-dimensional {H}olstein polaron with the {D}avydov
  ans\"atze.
\newblock {\em Phys. Rev. B}, 82:014305, Jul 2010.

\bibitem{xu-yan-JCP2010}
R-X. Xu, P.~Cui, X-Q. Li, Y.~Mo, and Y-J. Yan.
\newblock Exact quantum master equation via the calculus on path integrals.
\newblock {\em {J}. {C}hem. {P}hys.}, 122:041103, 2005.

\bibitem{Gelzinis2011}
Andrius Gelzinis, Darius Abramavicius, and Leonas Valkunas.
\newblock Non-markovian effects in time-resolved fluorescence spectrum of
  molecular aggregates: Tracing polaron formation.
\newblock {\em Phys. Rev. B}, 84:245430, Dec 2011.

\bibitem{Mukamel2000}
S.~Mukamel.
\newblock Multidimensional femtosecond correlation spectroscopies of electronic
  and vibrational excitations.
\newblock {\em Annu. Rev. Phys. Chem.}, 51:691--729, 2000.

\bibitem{jonasARPC2003}
D.~M. Jonas.
\newblock Two-dimensional femtosecond spectroscopy.
\newblock {\em Annu. Rev. Phys. Chem.}, 54:425--463, 2003.

\bibitem{Brixner2005}
T.~Brixner, J.~Stenger, H.~M. Vaswani, M.~Cho, R.~E. Blankenship, and G.~R.
  Fleming.
\newblock Two-dimensional spectroscopy of electronic couplings in
  photosynthesis.
\newblock {\em Nature}, 434(7033):625--628, 2005.

\bibitem{ZigmantasFlemingPNAS2006}
D.~Zigmantas, E.~L. Read, T.~Man\v{c}al, T.~Brixner, A.~T. Gardiner, R.~J.
  Cogdell, and G.~R. Fleming.
\newblock Two-dimensional electronic spectroscopy of the {B800-B820}
  light-harvesting complex.
\newblock {\em Proc. {N}atl. {A}cad. {S}ci. {USA}}, 103(34):12672--12677, 2006.

\bibitem{engel-nat2007}
G.~S. Engel, T.~R. Calhoun, E.~L. Read, T.~K. Ahn, T.~Man\v{c}al, Y.~C. Cheng,
  R.~E. Blankenship, and G.~R. Fleming.
\newblock Evidence for wavelike energy transfer through quantum coherence in
  photosynthetic systems.
\newblock {\em Nature}, 446:782, 2007.

\bibitem{Collini2009}
E.~Collini and G.~D. Scholes.
\newblock Coherent intrachain energy migration in a conjugated polymer at room
  temperature.
\newblock {\em Science}, 323(5912):369--373, 2009.

\bibitem{Sperling2010}
Jaroslaw Sperling, Alexandra Nemeth, J\"{u}rgen Hauer, Darius Abramavicius,
  Shaul Mukamel, Harald~F. Kauffmann, and Franz Milota.
\newblock Excitons and disorder in molecular nanotubes: A 2d electronic
  spectroscopy study and first comparison to a microscopic model.
\newblock {\em J.~{P}hys.~{C}hem.~{A}}, 114(32):8179--8189, 2010.

\bibitem{Turner2011}
Daniel~B. Turner, Krystyna~E. Wilk, Paul M.~G. Curmi, and Gregory~D. Scholes.
\newblock Comparison of electronic and vibrational coherence measured by
  two-dimensional electronic spectroscopy.
\newblock {\em {J}. {P}hys. {C}hem. {L}ett.}, 2(15):1904--1911, 2011.

\bibitem{Westenhoff2012JACS}
Sebastian Westenhoff, David Pale\u{c}ek, Petra Edlund, Philip Smith, and
  Donatas Zigmantas.
\newblock Coherent picosecond exciton dynamics in a photosynthetic reaction
  center.
\newblock {\em J. Am. Chem. Soc.}, 134(40):16484--16487, 2012.

\bibitem{Panitchayangkoon2011}
Gitt Panitchayangkoon, Dmitri~V. Voronine, Darius Abramavicius, Justin~R.
  Caram, Nicholas H.~C. Lewis, Shaul Mukamel, and Gregory~S. Engel.
\newblock Direct evidence of quantum transport in photosynthetic
  light-harvesting complexes.
\newblock {\em Proc. Natl. Acad. Sci. USA}, 108(52):20908--20912, 2011.

\bibitem{Christensson_JPCB2012}
Niklas Christensson, Harald~F. Kauffmann, T{\~o}nu Pullerits, and Tom{\`a}\v{s}
  Man\v{c}al.
\newblock Origin of long-lived coherences in light-harvesting complexes.
\newblock {\em {J}. {P}hys. {C}hem. {B}}, 116(25):7449--7454, 2012.

\bibitem{Butkus-Zigmantas-Abramavicius-Valkunas-CPL2012}
V.~Butkus, D.~Zigmantas, L.~Valkunas, and D.~Abramavicius.
\newblock Vibrational vs. electronic coherences in 2{D} spectrum of molecular
  systems.
\newblock {\em {C}hem. {P}hys. {L}ett.}, 545:40--43, 2012.

\bibitem{Jonas_PNAS2012}
Vivek Tiwari, William~K. Peters, and David~M. Jonas.
\newblock Electronic resonance with anticorrelated pigment vibrations drives
  photosynthetic energy transfer outside the adiabatic framework.
\newblock {\em Proc. {N}atl. {A}cad. {S}ci. {USA}}, 110(4):1203--1208, 2013.

\bibitem{Mancal_JPC2012}
Tom{\'a}\v{s} Man\v{c}al, Niklas Christensson, Vladim{\'i}r Luke\v{s}, Franz
  Milota, Oliver Bixner, Harald~F. Kauffmann, and J{\"u}rgen Hauer.
\newblock System-dependent signatures of electronic and vibrational coherences
  in electronic two-dimensional spectra.
\newblock {\em {J}. {P}hys. {C}hem. {L}ett.}, 3(11):1497--1502, 2012.

\bibitem{Chin2013}
A.~W. Chin, J.~Prior, R.~Rosenbach, F.~Caycedo-Soler, S.~F. Huelga, and M.~B.
  Plenio.
\newblock The role of non-equilibrium vibrational structures in electronic
  coherence and recoherence in pigment-protein complexes.
\newblock {\em {N}ature {P}hys.}, 9(2):113--118, February 2013.

\bibitem{Polyutov2012}
Sergey Polyutov, Oliver K{\"u}hn, and T{\~o}nu Pullerits.
\newblock Exciton-vibrational coupling in molecular aggregates: Electronic
  versus vibronic dimer.
\newblock {\em {C}hem. {P}hys.}, 394(1):21--28, 2012.

\bibitem{Abramavicius2010}
D.~Abramavicius, V.~Butkus, J.~Bujokas, and L.~Valkunas.
\newblock Manipulation of two-dimensional spectra of excitonically coupled
  molecules by narrow-bandwidth laser pulses.
\newblock {\em Chem.~{P}hys.}, 372(1-3):22--32, 2010.

\bibitem{Abramavicius2007}
D.~Abramavicius, L.~Valkunas, and S.~Mukamel.
\newblock Transport and correlated fluctuations in the nonlinear optical
  response of excitons.
\newblock {\em Europhys. Lett.}, 80(1):17005, 2007.

\bibitem{Turner2012}
Daniel~B Turner, Rayomond Dinshaw, Kyung-Koo Lee, Michael Belsley, Krystyna~E
  Wilk, Paul~MG Curmi, and Gregory Scholes.
\newblock Quantitative investigations of quantum coherence for a
  light-harvesting protein at conditions simulating photosynthesis.
\newblock {\em Phys. Chem. Chem. Phys.}, 14:4857--4874, 2012.

\bibitem{Calhoun2009}
T.~R. Calhoun, N.~S. Ginsberg, G.~S. Schlau-Cohen, Y-C. Cheng, M.~Ballottari,
  R.~Bassi, and G.~R. Fleming.
\newblock Quantum coherence enabled determination of the energy landscape in
  light-harvesting complex {II}.
\newblock {\em {J}. {P}hys. {C}hem. {B}}, 113:16291--16295, 2009.

\bibitem{Butkus-Abramavicius-Valkunas-JCP2012}
Vytautas Butkus, Leonas Valkunas, and Darius Abramavicius.
\newblock Molecular vibrations-induced quantum beats in two-dimensional
  electronic spectroscopy.
\newblock {\em J.~Chem.~Phys.}, 137(4):044513, 2012.

\bibitem{Chenu2013}
Aur{\'e}lia Chenu, Niklas Christensson, Harald~F Kauffmann, and Tom{\'a}{\v{s}}
  Man{\v{c}}al.
\newblock Enhancement of vibronic and ground-state vibrational coherences in 2d
  spectra of photosynthetic complexes.
\newblock {\em Sci. Rep.}, 3:2029, 2013.

\bibitem{Schlau-Cohen2009}
Gabriela~S. Schlau-Cohen, Tessa~R. Calhoun, Naomi~S. Ginsberg, Elizabeth~L.
  Read, Matteo Ballottari, Roberto Bassi, Rienk van Grondelle, and Graham~R.
  Fleming.
\newblock Pathways of energy flow in lhcii from two-dimensional electronic
  spectroscopy.
\newblock {\em {J}. {P}hys. {C}hem. {B}}, 113(46):15352--15363, 2009.

\end{thebibliography}
\end{document}